
\overfullrule=0pt
\magnification=\magstep1
\input vanilla.sty
\input abb.tex
\title Invariants of 3-Manifolds Derived from\\ Finite Dimensional Hopf
Algebras\endtitle
\author {\rm by}\\
Louis H. Kauffman\\
{\rm and}\\
David E. Radford\\
\\
{\rm Department of Mathematics, Statistics}\\
{\rm and Computer Science  (M/C 249)}\\
{\rm University of Illinois at Chicago}\\
{\rm 851 South Morgan Street}\\
{\rm Chicago, Illinois  60607-7045}\endauthor

\sub{Abstract}  This paper studies invariants of 3-manifolds derived
from certain finite dimensional Hopf algebras via regular isotopy
invariants of unoriented links in the blackboard framing.  The
invariants
are based on right integrals for these Hopf algebras.  It is shown that
the resulting class of invariants is definitely distinct from the class
of Witten-Reshetikhin-Turaev invariants.  The invariant associated with
the quantum double of a finite group $G$ is treated in this context, and
is shown to count the number of homomorphisms of the fundamental group
of the 3-manifold to the given finite group $G$.
\mbk
\sub{Introduction}
\mbk
The purpose of this paper is to indicate a method of defining invariants
of 3-manifolds intrinsically in terms of right integrals on certain Hopf
algebras.  We call such an invariant a {\it Hennings invariant}  [5], as
Hennings was the first person to point out that invariants could be
defined in this way.
\mbk
Part of the motivation of this work is to understand the relationship of
the Hennings invariant with the invariants of Reshetikhin-Turaev [19]
defined in terms of surgery on framed links and the representation
theory
of the Hopf algebras.  In a sense, the Hennings invariants are closer in
spirit to the original Witten definition [21] in terms of functional
integration, since that approach gave a direct definition of the
invariant in terms of geometrical data (albeit in the context of quantum
field theory and at the physical level of rigor).  Here we consider
invariants defined directly in terms of the geometry of link
diagrams and the global structure of the algebra.
\mbk
Hennings invariants were originally defined using oriented links.  It is
not necessary to use invariants that are dependent on link orientation
to define 3-manifold invariants via surgery and Kirby calculus.  For
that reason, the invariants discussed in this paper are formulated for
unoriented links.  This results in a simplification and conceptual
clarification of the relationship of Hopf algebras and link invariants.
The practical benefit is a simplified algorithmic structure for the
calculation or reasoning about the invariants.  {\it Further reference
to invariants of 3-manifolds in this paper will, unless otherwise
specified, be to this version of the Hennings invariant for unoriented
links.}
\mbk
We show that invariants defined in terms of right integrals, as
considered in this paper, are definitely distinct from the invariants of
Reshetikhin and Turaev.  We show that our invariant is non-trivial for
the quantum group $U_q(sl_2)$' when $q$ is an fourth root of unity.  The
Reshetikhin-Turaev invariant is trivial at this quantum group and root
of unity.  The non-triviality of our invariant is exhibited by showing
that it distinguishes all the Lens spaces L(n,1) from one another.
This proves that there is non-trivial topological information in the
non-semisimplicity of $U_q(sl_2)$'.
\mbk
It is clear that our result opens the door to a class of new invariants
of 3-manifolds derived from quantum groups.  This paper concentrates on
setting out the general properties of the invariant and the specific
calculations for Lens spaces alluded to above.
\mbk
We also include a treatment of the invariant derived from the Drinfeld
double associated with a finite group $G$.  We show that this invariant,
defined by a right integral, counts the homomorphisms of the fundamental
group of the 3-manifold to the group $G$.  This is the expected result
and this context provides a neat proof of it.  In a sequel to the
present paper, we shall give other calculations and examples.
\mbk
The paper is organized as follows:  Section 1 recalls Hopf algebras,
quasi-triangular Hopf algebras and ribbon Hopf algebras.  Section 2
discusses the conceptual setting of the invariant.  This involves a
summation over labellings of the link diagram by elements of the Hopf
algebra.  We work in a category that allows immersed diagrams so that
the special grouplike element in the Hopf algebra and the ribbon element
in the Hopf algebra both have diagrammatic interpretations.  A trace
function on the Hopf algebra that is invariant under the antipode is
shown to yield a link invariant.  Section 3 discusses the categorical
framework for the constructions in Section 2.  In Section 4, we show
that traces of the kind discussed in Section 2 are constructed from
right integrals in many cases and that under suitable conditions these
traces yield invariants of the 3-manifolds obtained by surgery on the
links.  Section 5 gives the promised application to $U_q(sl_2)$'.
Section 6 gives the application to the Drinfeld double associated with
a finite group.
\mbk
\sub{Acknowledgement}  L. Kauffman thanks N. Reshetikhin for helpful
conversations, the Department of Mathematics of the University College
of
Swansea, Swansea, Wales for hospitality during the preparation of this
paper, and the National Science Foundation for support of this research
under grant number DMS-9205277.  D. Radford thanks the National Science
Foundation for support of this research under grant numbers DMS-910622
and DMS-9308106.
\vfill\eject\noi
{\bf I. Algebra}
\mbk
Recall that a {\it Hopf algebra} $A$ [20] is a bialgebra over a
commutative ring $k$ that has associative multiplication, coassociative
comultiplication and is equipped with a counit, a unit and an antipode.
The ring $k$ is usually taken to be a field.
\sbk
In order to be an algebra, $A$ needs a multiplication $m:A
\otimes
A \raw A$.  The associative law for $m$ is expressed by the
equation $m(m \otimes 1) = m(1 \otimes m)$ where 1 denotes the
identity map on $A$.
\mbk
In order to be a bialgebra, an algebra needs a coproduct
$\Delta:A \raw  A \otimes A$.  The coproduct is a map of
algebras, and is regarded as the dual of multiplication.
The comultiplication $\Delta$ is coassociative.  Coassociativity of $\Delta$ is
the equation $(\Delta \otimes 1) \Delta = (1 \otimes \Delta) \Delta$
where 1 denotes the identity map on $A$.
\mbk
The unit is a mapping from $k$ to $A$ taking 1 in $k$ to 1 in $A$,
and thereby defining an action of $k$ on $A$.  It will be convenient to
just identify the units in $k$ and in $A$, and to ignore the name of the
map that gives the unit.
\mbk
The counit is an algebra mapping from $A$ to $k$ denoted by
$E:A \raw k$.  The following formulas for the counit dualize
the structure inherent in the unit:  $(E \otimes 1) \Delta = 1 =
(1 \otimes E) \Delta$.  Here the 1 denotes the identity map on $A$.
\mbk
It is convenient to write formally $\Delta(x) = \Sigma
x_{(1)} \otimes x_{(2)} \in A \otimes A$ to indicate the
decomposition
of the coproduct of $x$ into a sum of first and second factors in the
two-fold tensor product of $A$ with itself.  We shall further adopt the
summation convention which abbreviates $\Sigma x_{(1)}  \otimes x_{(2)}$
to just $x_{(1)} \otimes x_{(2)}$.  Thus, we write
$\Delta(x) = x_{(1)} \otimes x_{(2)}$.
\mbk
The antipode is a mapping s:$A \raw  A$ satisfying the
equations $m(1 \otimes s) \Delta (x) = E(x)1$ and $m(s \otimes 1)\Delta
(x)
= E(x)1$, where 1 on the right hand side of these equations denotes the
unit of $k$ as identified with the unit of $A$.  It is a consequence of
this definition that $s(xy) = s(y)s(x)$ for all $x$ and $y$ in $A$.
\mbk
A {\it quasitriangular Hopf algebra} $A$ [3] is a Hopf algebra with an
element $\rho \in A \otimes A$ satisfying the following
equations:
\mbk
1) $ \rho\Delta = \Delta'\rho$, where $\Delta'$ is the composition of
$\Delta$ with the map on $A \otimes A$ that switches the two factors.
\mbk
2) $ \rho_{12}\rho_{13} = (1 \otimes \Delta) \rho,\quad
\rho_{13}\rho_{23} = (\Delta \otimes 1)\rho$.
\mbk
These conditions imply that $\rho$ has an inverse, and that
$\rho^{-1} = (1 \otimes s^{-1}) \rho = (s \otimes 1) \rho$.
\mbk
It follows easily from the axioms of the quasitriangular Hopf algebra
that $\rho$ satisfies the Yang-Baxter equation
$$\rho_{12}\rho_{13}\rho_{23} = \rho_{23}\rho_{13}\rho_{12}.$$

A less obvious fact about quasitriangular Hopf algebras is that there
exists an element $u$ such that $u$ is invertible and $s^2(x) =
uxu^{-1}$
for all $x$ in $A$.  In fact, we may take $u = \Sigma s(e')e$ where
$\rho = \Sigma e \otimes e'$.
\mbk
An element $G$ in a Hopf algebra is said to be {\it grouplike} if
$\Delta(G) = G  \otimes G$ and $E(G) = 1$  (from which it follows that
$G$
is invertible and $s(G) = G^{-1})$.  A quasitriangular Hopf algebra is
said to be a {\it ribbon Hopf algebra} [18], [9]  if there exists a
grouplike element $G$ such that (with $u$ as in the previous paragraph)
$v = G^{-1}u$ is in the center of $A$ and $s(u) = G^{-1}uG^{-1}$.    We
call $G$ a {\it special} grouplike element of $A$.
\mbk
Since $v=G^{-1}u$ is central, $vx=xv$ for all $x$ in $A$.  Therefore
$G^{-1}ux = xG^{-1}u$, whence $s^2(x) = uxu^{-1} = GxG^{-1}$.  Thus
$s^2(x) = GxG^{-1}$ for all $x$ in $A$.
Similarly, $s(v) = s(G^{-1}u) = s(u)s(G^{-1}) = G^{-1}uG^{-1}G =G^{-1}u
= v$. Thus the square
of the antipode is represented by conjugation by the special grouplike
element in a ribbon Hopf algebra, and the central element $v = G^{-1}u$
is invariant under the antipode.
\vskip2em\noi
{\bf II.  Diagrammatic Geometry and the Trace}
\mbk
\noi
A function $tr:A \rightarrow K$  from the Hopf algebra to the base ring
$k$ is said to be a {\it trace} if
$$
\te{tr}(xy)  = \te{tr}(yx)$$
and
$$\te{tr}(s(x)) = \te{tr}(x)$$
for all $x$ and $y$ $\epsilon$ $A$.  In this section, we describe how
a trace function on a ribbon Hopf algebra yields an invariant, $TR(K)$,
of regular isotopy of knots and links  [6], [7].
\mbk
The link diagram is arranged with respect to a vertical direction so
that the crossings form the two types indicated below, and so that other
than the crossings the only critical points of the height function are
maxima and minima.  Each crossing is decorated with elements of the Hopf
algebra as shown below.  (Here $\rho = \Sigma e \otimes e$' is the
Yang-Baxter element in $A \otimes A$, and $s$ denotes the antipode.)
\vskip12em
It is implicit in this formalism that there is a summation over all the
pairs $e,e'$ for each Yang-Baxter element.
Hopf algebra elements may be moved across maxima or minima at the
expense of application of the antipode.  That is, if a Hopf algebra
element is moved across a maximum or minimum, then it is replaced by the
application of the antipode to that element if the motion is
anti-clockwise.  If the motion is clockwise, then the inverse of
\vfill\eject
\vskip3em
the antipode is applied to the element.  See the diagram below.
\vskip10em
The link diagram is subject to deformations that generate regular
isotopy [8].  Since the diagram is presented with respect to a choice of
vertical direction (discriminating the maxima, minima and crossing
types), regular isotopy is generated by a set of moves that include the
cancellation of adjacent pairs of maxima and minima and the switching of
an arc across a maximum or minimum.  The full set of moves is shown in
Figure 1.  We have labelled these moves as
\sbk
\item { --}
\item { --}    (cancellation of maxima and minima)
\item {II}     (cancellation of opposite crossings)
\item {III}    (braiding)
\item {IV}     (switching)
\item {IV'}    (twist of crossings).
\mbk
IV' is equivalent to IV  in the presence of the cancellation of maxima
and minima.  These moves generate regular isotopy for diagrams arranged
with respect to a vertical direction.
\vskip22em
\centerline{\bf Figure 1}
\vfill\eject
\vskip3em
\sub {\bf Remark}  The symbol \qquad is used to denote the
replacement of one
figure by an equivalent figure.  We shall sometimes use an equals sign
$(=)$ to perform the same purpose.  The symbol \qquad  or
\qquad will be used to indicate a
correspondence. For
example, a link diagram corresponds to the diagram obtained from it by
decoration with elements of the Hopf algebra.
\mbk
An invariant of regular isotopy  must remain unchanged by the moves
shown in \break\noi Figure~1.  The simplest move is the
cancellation of a pair consisting of a maximum and a minimum.
\vskip10em
This pair cancellation gives a reformulation of the slide rule for the
antipode:  The antipode is accomplished by ``composition with a maximum
and a minimum''.
\vskip15em
Note also that once the crossings of a link diagram have been labelled
with elements of the Hopf algebra, the resulting diagram is depicted as
a labelled immersion of a curve or curves in the plane.  This is quite
natural since the translation from algebraic braiding element to
knot-theoretic braiding element is accomplished via the composition with
a transposition, and the simplest diagrammatic representation of a
transposition is the crossing of two arcs in the plane.
\vskip6em
These immersions can be deformed up to regular homotopy that respects the
given vertical direction.  In other words, one can perform the projected
forms of the moves of
Figure 1 if algebra is present on the lines then
the following extra move is added (sliding an external line past an
algebra element).
\vskip1em \noi
V (slide rule)
\vskip8em
Since algebra elements are configured with respect to the vertical
direction, we do not allow the cancellation of a maximum and a minimum
that have an algebra element between them.  This allows the
representation of the antipode as described above.
It is now easy to check the twist relation (IV') for crossings:
\vskip15em
With these conventions, the square of the antipode is equivalently
diagrammed as a ``composition with two curls'' as shown below:
\vfill\eject
\vskip3em
These curls are identified with the special grouplike element $G$ and
$G^{-1}$ in the Hopf algebra.
\vskip10em
Thus the diagram for the square of the antipode represents directly the
formula $s^2(x) = GxG^{-1}$.

Note that this identification is compatible with the rule $s(G) =
G^{-1}$.
\vskip8em
Along a vertical line, algebra elements combine by multiplication.
\vskip8em
The product in the Hopf algebra corresponds to the multiplication of
single strand tangles.  (A single strand tangle is a bit of link diagram
with two free ends arranged with respect to the vertical so that one end
is down and the other end is up.  Tangles are multiplied by attaching
the down end of one tangle to the top end of the other).
\vskip12em
The coproduct $\Delta:A {\rightarrow} A{\otimes}A$ in the Hopf algebra
corresponds to a mapping on tangles $\Delta:T^{(1)} \rightarrow
T^{(2)}$ from single strand tangles to double stand tangles obtained by
forming the parallel (two strand) cable of the given tangle.  The
tangles in question can be immersions.
\vfill\eject
For example, we see that the formula $\Delta(G) = G {\otimes} G$
corresponds to the regular isotopy shown below.
\vskip9em
In this way knots on a line can be resolved into algebra elements.  For
example, the twist shown below is equivalent to the ribbon element $v$.
Note how the factorization of $v$ into a product of $G^{-1}$ and $u =
{\Sigma}s(e ')e$ is related to the slide convention for the antipode
(In the diagrammatic calculation shown below, we use the fact that
$(s{\otimes}s)\rho = \rho.$)
\vskip18em
Note also that diagrammatically $s(v) = v$ corresponds to the
identification of twists shown below.
\vskip10em
When this identification is added to regular isotopy, the twists catalog
only the framing, and the equivalence relation on the link diagrams is
equivalent to ambient isotopy of framed links.
\vfill\eject
Here is an example of the translation of the single strand trefoil
tangle into algebra:
\vskip22em
Finally, returning to the diagrammatic coproduct, we see the
interpretation of the following formula of Drinfeld
$\Delta(u) = \rho_{21}\rho_{12}(u{\otimes}u)$ :
\vskip22em
In general, if $T$ is a single strand tangle, and $F(T)$ is the
corresponding element in the Hopf algebra $A$ that is determined by our
correspondence, then $F \Delta(T) = \Delta(F(T))$ where the first
$\Delta$ is the diagrammatic coproduct and the second $\Delta$ is the
algebraic coproduct.

This fact follows from the axioms for a quasi-triangular Hopf algebra in
conjunction with our diagrammatic conventions.

\sub{Definition and Computation of $TR(K)$}
\mbk
Suppose that $tr:A \rightarrow k$ is a trace function.  That is, $tr$ is
a linear function satisfying $\te{tr}(xy) = \te{tr}(yx)$ and
$\te{tr}(s(x)) = \te{tr}(x).$
To define the trace $TR(K)$ for a knot diagram $K$, slide all of the
algebra into one vertical portion of the diagram.  Amalgamate this
algebraic expression according to the rule for multiplying algebra
elements on the diagram, as we have done above.  Call this localized
algebra element  $w$.  It is a sum of products, and can be formally
represented as a product where it is understood that there is a sum over
all pairs of the type $e,e $'.

Let  $d$ be the Whitney degree of the flat diagram for $K$ that is
obtained by traversing $K$ upward from the vertical portion where the
algebra has been concentrated.  The Whitney degree is the total turn of
the tangent vector to the curve as one traverses it in the given
direction.  For example:
\vskip7em
Define $TR(K)$ by the formula $TR(K) = tr(wG ^d)$.
Note that $w$ is itself a summation over all the pairs $x,x $'
corresponding to Yang-Baxter elements on the diagram.  $TR(K)$ {\it
defines a regular isotopy invariant of unoriented knots}.  (The proof is
primarily a matter of checking that $TR(K)$ is independent of the place
where we concentrate the algebra.  This reduces to checking the
independence in the case where the concentration is moved around a
maximum or a minimum.  See example 2 below; and for a complete proof see
Theorem 5.1 of [7].)
In order to define an invariant of unoriented links, concentrate the
algebra for each component of the link, and define
$$TR(K) = tr(w_1G^{d_1})tr(w_2G^{d_2})tr(w_3G^{d_3})  ... tr(w_n
G^{d_n})$$
where the labels $1,2,...,n$  refer to the components' of the link, and
the implicit summation is the sum over all the pairs $x,x'$ in these
words.  The elements $w_1, ..., w_n$ are the algebra concentrations for
each link component, and the degrees  $d_1, ..., d_n$ are the Whitney
degrees of the components of the link.
\vfill\eject
\sub{Example 1}
\vskip12em
Here $u = \Sigma s(e ')e$ and $G$ is the special grouplike element.
Thus $TR(K) = tr(uG^0) = tr(u)$.

\sub {Example 2}  This example points out how the $TR(K)$ is invariant
under algebra slides:
\vskip9em
$tr(s(x)G) = tr(s(s(x)G)) = tr(G^{-1}s^2(x)) = tr(G^{-1}GxG^{-1}) =
tr(xG^{-1})$.
\vskip1em
\sub {Example 3}  Here is the form of calculation for a link.
\vskip10em
$TR(L) = \Sigma tr(f 'eG^{-1})tr(fe 'G)$.

If $\rho =\sum_{i=1}^nx_i{\otimes}y_i$, then $TR(L) = \sum_{i=1}^n
\sum_{j=1}^n tr(y_jx_iG^{-1})tr(x_jy_iG)$

This is how the regular isotopy invariant of the link would look as a
specific sum of traces of algebra elements.
\vskip2em\noi
{\bf III.  Categories}
\mbk
The description of invariants given in Section II may seem a bit ad hoc.
The purpose of this section is to give a quick description of our
methods in categorical terms.  In these terms, we see that we have
functors between categories of diagrams and categories naturally
associated with Hopf algebras.

A link diagram, with or without free ends, that is arranged with respect
to a vertical direction can be regarded as a morphism in a tensor
category with two generating objects $V$ and $k$.  We think of $V$ as
the analog of a vector space or module over $k$.  Thus $k \otimes k =
k$, $k \otimes V = V \otimes k = V$, and the products $V$,
$V \otimes V$,
$V \otimes V \otimes V$, ... are all distinct.  Maxima are morphism from
$V \otimes V$ to $k$, and denoted $\cap: V \otimes V \rightarrow k$.
Minima are morphisms from $k$ to $V \otimes V$, and are denoted
$\cup: k \rightarrow V \otimes V$.  The two types of crossing yield
morphisms $R,\bar R: V \otimes V \rightarrow V \otimes V$ from
$V \otimes V$ to itself.  See the diagrams below.
\vskip22em
Composition of morphisms corresponds to tying input and output lines
together in these diagrams.  In this way, each diagram with out any
tangle lines becomes a morphism from $k$ to $k$.  We call such a diagram
a {\it closed diagram}. and we shall call the corresponding morphism
from $k$ to $k$ a {\it closed morphism}.  The trace function described
in the last section is designed to assign a specific element in a
commutative ring (e.g. the complex number) to such morphism from $k$ to
$k$.

The morphism in this regular isotopy category of link diagrams (denoted
REG) are assumed to satisfy those relations that correspond to regular
isotopy of links with respect to a vertical direction.  See Figure 1.

We also use the category of FLAT diagrams.  In these, the crossing has
no distinction between over and under, but there is still a
(transposition) morphism $P: V \otimes V \rightarrow V \otimes V$
$(P(a \otimes b) = b \otimes a)$ corresponding to a crossing.  The
axioms
for the category of FLAT diagrams are exactly the same in form as the
axioms for the category of REG diagrams,
\vfill\eject\noi
except that there are no
restrictions about the types of crossing since there is only one type of
crossing:
\vskip6em
We know from the Whitney-Graustein theorem that FLAT has a particularly
simple structure of closed morphisms.  In terms of diagrams any such
morphism has the form shown below.
\vskip17em
It is for this reason that we have singled out the morphisms $G$ and
$G^{-1}$ in the previous section.  We use the identification
$V = V \otimes k$ and take $G =
(1 \otimes \cap) (P \otimes 1)(1 \otimes 1)(1 \otimes \cup)$ as a
morphism
from $V$ to $V$.  As we have already remarked, $G$ corresponds to
a flat curl, and in the context of labelling the diagram with elements
of a quantum group, $G$ corresponds to a grouplike element.
\vskip17em
For both categories REG and FLAT, we have functors
$\Delta:REG \rightarrow REG{\otimes}REG$ and $\Delta:FLAT \rightarrow
FLAT \otimes FLAT$.  On objects $W$, $\Delta(W) = W \otimes W$.
On morphisms, $\Delta(T)$ is obtained by taking the 2-fold parallel cable
of each component arc in $T$.

In order to describe labelled diagrams in categorical language, we begin
with a Hopf algebra $A$, and associate to  $A$ a category
$C(A)$.  The category $C(A)$ has two generating objects $V$ and
$k$ with the same properties as the $V$ and $k$ for REG and FLAT.
However, $k$ is taken to be identical with the ground ring for the Hopf
algebra.  {\it The morphisms of C(A) are the elements of A}.  Each
morphism takes $V$ to $V$ and composition of elements corresponds to
multiplication in $A$.

The labelled diagrams of simplest form
\vskip10em \noi
exhibit elements of $A$ in the role of morphisms in $C(A)$.

We then define categories $C(A) \ast REG$ and $C(A) \ast FLAT$, by
adding
the morphisms $\cup$,$\cap$,$R$,$\ol R$ or $P$ to $C(A)$ in the form of
diagrams
as we have already described them in the previous section.  Note how
$\cup$ and $\cap$ are defined to carry
the structure of the antipode of $A$.

One then sees that we have, in the previous section, defined a functor
from REG to $C(A)\ast FLAT$, and then by ``moving elements around the
diagram'', shown how to reduce closed morphisms in C(A)*FLAT to a sum of
formal traces of products of elements in $A$.  This functor $F:REG
\rightarrow C(A)*$FLAT is the source of the invariants.

This categorical description of the invariants is compatible with a
representation of the Hopf algebra as linear mappings of a vector space
$V$ over a field $k$.  In that case, the cup and the cap must actually
be mappings between $V \otimes V$ and $k$.  The abstract compatibility
relations between cup, cap and antipode then translate into specific
conditions that allow the invariant to be computed either by direct
composition of linear maps \quad (The value of the invariant on a
closed morphism $\mu:k \rightarrow k$ is $ \mu(1)$.) or by first
concentrating the algebra as in section 2.  In section 6, this method is
used to calculate the invariant associated with a finite group.
\vskip2em\noi
{\bf IV. Invariants of 3-manifolds}
\mbk
The structure we have built so far can be used to construct invariants
of 3-manifolds presented in terms of surgery on framed links.  We sketch
here our technique that simplifies an approach to 3-manifold invariants
of Mark Hennings [5].

Recall that an element $\lambda$ of the dual algebra $A^*$ is said to be
a {\it right integral\/} if \hfill\break\noi  $\lambda(x)1 =
m(\lambda \otimes 1)(\Delta(x))$ for all $x$ in $A$.  For a unimodular
[12],[15] finite dimensional\hfill\break\noi ribbon Hopf algebra $A$
over a field $k$ there
is a right integral $\lambda$ satisfying the following properties for
all $x$ and $y$ in $A$:
\vfill\eject
\item {0)}   $\lambda$ is unique up to scalar multiplication.
\item {1)}   $\lambda(xy) = \lambda(s^2(y)x)$.
\item {2)}   $\lambda(gx) = \lambda(s(x))$ where $g=G^2$,
and $G$ is the special grouplike element for the ribbon element $v=G^{-1}u$.
\mbk
Given the existence of this $\lambda$, define a functional $tr:A
\rightarrow k$ by the formula \hfill \break \noi $tr(x) = \lambda(Gx)$.
\mbk
\sub{Theorem}  With $tr$ defined as above, then
\item {1)}  $tr(xy) = tr(yx)$ for all $x,y$ in $A$.
\item {2)}  $tr(s(x)) = tr(x)$ for all $x$ in $A$.
\item {3)}  $[m(tr \otimes 1)(\Delta(u^{-1}))]u = \lambda(v^{-1})v$
where $v=G^{-1}u$ is the ribbon element.
\mbk
\sub{Proof}  The proof is a direct consequence of the properties 1)
and 2) of $\lambda$.
Thus
$$
tr(xy) = \lambda(Gxy) = \lambda(s^2(y)Gx) =
\lambda(GyG^{-1}Gx) = \lambda(Gyx) = tr(yx),$$ and
$$\align tr(s(x)) &=
\lambda(Gs(x)) = \lambda(gG^{-1}s(x)) = \lambda(s(G^{-1}s(x))) =
\lambda(s^2(x)s(G^{-1}))\\ & = \lambda(s^2(x)G) = \lambda(GxG^{-1}G) =
\lambda(Gx) = tr(x).\endalign$$
Finally,
$$\align
[m(tr \otimes 1)(\Delta(u^{-1}))]u &= G^{-1}
[m(\lambda\cdot G \otimes G)(\Delta(u^{-1}))]u\\ & = [m(\lambda
\otimes 1)(\Delta(Gu^{-1}))]G^{-1}u = \lambda(Gu^{-1})G^{-1}u =
\lambda(v^{-1})v.\endalign $$
This completes the proof.\hfill$Q.E.D.$
\vskip2em
The upshot of this Theorem is that for a unimodular finite dimensional
Hopf algebra there is a natural trace defined via the existent right
integral.  Remarkably, this trace is just designed by property 3) of the
Theorem to behave well with respect to the Kirby move.  The Kirby move
is the basic transformation on framed links that leaves the
corresponding 3-manifold obtained by framed surgery unchanged.  See
[11], [19].  This means that a suitably normalized version of this trace
on framed links gives an invariant of 3-manifolds.  To see how this
works, here is a sample Kirby move
\vskip12em
The cable going through the loop can have any number of strands.  The
loop has one strand and the framing as indicated.  The replacement on
the right hand side puts a 360 degree twist in the cable with blackboard
framing as shown above.  Here we calculate the case of a single strand
cable:
\vskip12em
The diagram shows that the trace contribution is \quad (with implicit
summation on the repeated primed and unprimed pairs of Yang-Baxter
elements)
$$\align
tr(f'v^{-1}eG^{-1})fe' &= tr(f'ev^{-1}G^{-1})fe'
= tr(f'eu^{-1})fe'\\
&= [m(tr \otimes 1)(f'eu^{-1} \otimes fe'u^{-1})]u\\ & = [m(tr \otimes
1)
(\rho_{21}\rho_{12}(u^{-1} \otimes u^{-1}))]u\\ &= [m(tr \otimes
1)(\Delta (u^{-1}))]u = \lambda(v^{-1})v.\endalign$$

$$(\Delta(u^{-1}) =\rho_{21}\rho_{12} (u^{-1} \otimes u^{-1}))$$

It follows from this calculation that the evaluation of the left hand
picture in the Kirby move is $\lambda(v^{-1})$ times the evaluation of
the right hand picture.  The corresponding result for an n-strand cable
is obtained by applying the coproduct to the equation above, and using
the functoriality of the coproduct with respect to tangles and tensor
powers of the Hopf algebra.  The diagram below illustrates this point
for the case of a 2-strand cable.
\vskip11em
Thus a proper normalization of $TR(K)$ gives an invariant of the
3-manifold obtained by framed surgery on $K$.  More precisely,
(assuming that $\lambda(v)$ and $\lambda(v^{-1})$ are non-zero) let
\mbk
$ INV(K)= {[\lambda(v)\lambda(v^{-1})]^{-c(K)/2}[\lambda(v)/
\lambda(v^{-1}]^{-\sigma(K)/2}} TR(K)$
\mbk \noi
where $c(K)$ denotes the number of components of $K$, and $\sigma(K)$
denotes
the signature of the matrix of linking numbers of the components of $K$
(with framing numbers on the diagonal), then  $INV(K)$ is an invariant
of the 3-manifold obtained by doing framed surgery on $K$ in the
blackboard framing.  This is our reconstruction of Hennings' invariant
[5] in an intrinsically unoriented context.
\vskip2em
\sub{Remark 1} It is easy to prove directly the full handle-sliding
invariance of $TR(K)$. This follows directly from the fact that
$\lba (x)1=m(\lba\otimes 1)\bigl(\Dta(x)\bigr)$, the definition of
$TR(K)$ in
terms of $\lba$, and the functoriality of $\Dta$ expressed by the
equation
$F(\Dta(T))=\Dta F(T)$ where $T$ is a 1-1 tangle, $\Dta(T)$ is the 2-2
tangle obtained by making a parallel cable of $T$ and $F$ is the functor
described in section III.  The functor $F$ takes 1-1 tangles to the Hopf
algebra $A$ and can itself be studied as a source of invariants of 1-1
tangles. This approach will be taken in a sequel to the present paper.
\bbk
\sub{Remark 2} Note that since $S^1\times S^2$ is obtained by surgery on an
unframed circle, it follows at once from our definitions that
$INV (S^1\times S^2)=0$. This leads to questions about the relationship
of this invariant to properties of topological quantum field theories.  We plan
to continue this part of the investigation in a separate paper.
\vskip2em\noi
{\bf V.} ${\bold U}_{\bold q}({\bold s}{\bold l}_{\bold 2})'$
\mbk
The purpose of this section is to set up part of the general
calculations for $U_q(sl_2)'$, and in particular to calculate the
special case of the evaluation of the right integral on powers of the
ribbon element $v$ in the case $n=8$.  This will give us the result that
the invariant $INV(K)$ is distinct from the
Witten-Reshetikhin-Turaev invariant at this root of unity.

Recall the algebraic structure of $U_q(sl_2)'$.

Let $t$ be a primitive n-th root of unity, $q = t^2$, $m =
order(t^4)$.
Assume $m\ne 1$ (that is $n \ne 1,2,4)$.
The algebra has generators and relations as given below.
$$
\align
ae &=qea\\
af &=q^{-1}fa \\
a^n &= 1 \\
e^m &=0=f^m \\
{[e,f]} &= ef-fe = (a^2 - a^{-2})/(q - q^{-1})\endalign
$$
\mbk
The Yang-Baxter element is given by the formula below [11],[16].
\mbk
$$
R = \sum ^{m-1}_{v=0} \sum_{i,u \epsilon Z_n}
[(t^{-uv-i(u-v)-v}(q-q^{-1})^v)/(n(v)_q!)] f^va^i \otimes
e^va^{-u}.
$$
Here $(\ell)_q =(q^\ell-1)/(q-1)$ and $(\ell)_q! =
(\ell)_q(\ell-1)_q \lds (1)_q$.
\mbk
The coproduct is described by the formulas
$$
\align
\Delta a &= a \otimes a \\
\Delta x &= x \otimes a^{-1} + a \otimes x, x=e,f.
\endalign
$$
The counit is determined by the formulas
$$E(e) = E(f) = 0$$ and $$E(a) = 1.$$
It follows from the definition of the antipode $s$ that for
$x=e$ or $f$, \break\noi $0 = E(x)1 = m(s \otimes 1)
\Delta(x)=s(x)a^{-1}+s(a) x = s(x)a^{-1} + a^{-1}x$.
$(s(a) = a^{-1} \te{ since}\ph{a} \Delta(a) = a \otimes a.)$
This means $s(x) = -a^{-1}xa$, whence
$$s(e) = -q^{-1}e$$
and
$$s(f) = -qf.$$
The special grouplike element is  $G = a^{-2}$.

The special element $u$ such that $s^2(x)=uxu^{-1}$ for all $x$ is
given by the formula $u=\sum s(R^{(2)})R^{(1)}$. The next lemma gives a
specific formula for $u$.

\sub{Lemma 1} $u=\sum\limits^{m-1}_{v=0}\sum\limits_{i,j\in Z_n}
\bigl[(t^{j(i-v)-i^2-3v}(q^{-1}-q)^v)/\bigl(n(v)_q!\bigr)\bigr]a^je^vf^v$.

\sub{Proof} $u=\sum s(R^{(2)})R^{(1)}$
$$\align
&=\sum^{m-1}_{v=0} \sum_{i,u\in Z_n}
\bigl[(t^{-uv+i(u-v)-v}(q-q^{-1})^v)/(n(v)_q!)\bigr]s(e^va^{-u})f^va^i\\
&=\sum^{m-1}_{v=0}\sum_{i,u\in Z_n}
\bigl[(t^{-uv+i(u-v)-v}(q-q^{-1})^v)/(n(v)_q!)\bigr]a^u(-q^{-1})^ve^vf^va^i\\
&(q=t^2)\\
&=\sum^{m-1}_{v=0}\sum_{i,u\in Z_n}
\bigl[(t^{-uv+i(u-v)-3v}(q^{-1}-q)^v)/(n(v)_q!)\bigr]a^{u+i}e^vf^v\\
&(j=u+i \ \te{ so } \ i=j-u)\\
&=\sum^{m-1}_{v=0} \sum_{u,j\in Z_n}
\bigl[(t^{-uv+(j-u)(u-v)-3v}(q^{-1}-q)^v)/(n(v)_q!)\bigr]a^je^vf^v\\
&=\sum^{m-1}_{v=0} \sum_{u,j\in
Z_n}\bigl[(t^{j(u-v)-u^2-3v}(q^{-1}-q)^v)/(n(v)_q!)\bigr]a^je^vf^v.\endalign$$
Thus
$$u=\sum^{m-1}_{v=0} \sum_{i,j\in Z_n} \bigl[(t^{j(i-v)-i^2-3v}
(q^{-1}-q)^v)/(n(v)_q!)\bigr] a^j e^v f^v.$$\hfill$Q.E.D.$
\vskip2em
\sub{Change of Basis}
\mbk
We now make the following change of basis.  Replace $e$ by
$-(q-q^{-1})e$. Then
$$\align
ae &= qea\\
af &= q^{-1}fa\\
a^n &= 1\\
e^m &=0=f^m\\
[f,e] &= a^2-a^{-2}.\endalign$$
Note that in this basis the formula for $u$ becomes
$$u=\sum^{m-1}_{v=0} \sum_{i,j\in Z_n}
\bigl[(t^{j(i-v)-i^2-3v})/(n(v)_q!)\bigr]a^je^vf^v.$$
\vskip2em
\sub{Right Integral}
\mbk
A right integral $\lba$ for $A=U_q(sl_2)'$ is described as follows.
Consider the linear basis for $A$ given by the set
$\{a^ie^if^k\ph{0}|\ph{0}0\le
i<n,\ 0\le j, k<m\}$. Then $\lba(w)$ for $w\in A$ is the coefficient of
$a^{2(m-1)}e^{m-1}f^{m-1}$ in a writing of $w$ in this basis.  We
write \hfill\break\noi$\lba=\ol{a^{2(m-1)} e^{m-1} f^{m-1}}$ where the
bar over the
expression denotes the characteristic function of this element of the
algebra $A$. That this formula gives the right integral can be verified
by direct calculation [17].
\bbk
\sub{Orthogonal Idempotents}
\mbk
Let $\Lba_i=(1/n)\sum\limits_{j\in Z_n} t^{ij}a^j$. Then
$\Lba_i\Lba_j=\Lba_i\dta_{ij}$ where $\dta_{ij}$ is the Kronecker delta
and $1=\Lba_0+\Lba_1+\cdots +\Lba_{n-1}$.

Thus $\{\Lba_0, \Lba_1,\lds, \Lba_{n-1}\}$ form a set of orthogonal
idempotents for the group algebra $k[H]$ where $H=(a)=Z_n$.
\mbk\noi
{}From the relation
$$
\sum_{i\in Z_n}t^{ik}=
\left\{\aligned
n &\te{ if } k=0\\
0 &\te{ if } k\ne 0\endaligned\right.$$
for $k\in Z_n$, we have

\sub{Lemma 2} $a=\sum\limits_{i\in Z_n} t^{-i}\Lba_i$.

\sub{Proof}
$$\align
\sum_{i\in Z_n} t^{-i}\Lba_j &= \sum\limits_{i,j\in Z_n}
(t^{-i}t^{ij}/n)a^j=\sum_{j\in Z_n} \biggl(\sum_{i\in Z_n}
(t^{i(j-1)}/n)\biggr)a^j\\
&\\
&= (1/n)na=a.\endalign$$\hfill$Q.E.D.$
\vskip2em\noi
Hence
$$\align
u &=\sum^{m-1}_{v=0} \sum_{i\in Z_n} \biggl(\sum_{j\in
Z_n}\bigl[(t^{-i^2-3v})/\bigl((v)_q!\bigr)\bigr]\bigl[t^{j(i-v)}a^j/n
\bigr]\biggr) e^vf^v\\
&=\sum^{m-1}_{v=0} \biggl(\sum_{i\in Z_n}
(t^{-i^2-3v}/(v)_q!)\Lba_{i-v}\biggr)e^vf^v.\endalign$$

\sub{Lemma 3} $u= c\biggl(\sum\limits^{m-1}_{v=0}
\bigl[(t^{-3v-v^2})/(v)_q!\bigr]a^{2v}e^vf^v\biggr)$ where \ph{c}
$c=\sum\limits_{i\in Z_n} t^{-i^2}\Lba_i$.

\sub{Proof} As $a^{-2v}=\sum\limits_{i\in Z_n}
t^{2vi}\Lba_i=\sum\limits_{i\in Z_n} t^{2v(i-v)}\Lba_{i-v}$, we have
$$\align
&(\te{using } \Lba_i\Lba_j=\Lba_i\dta_{ij})\\
u &= \sum^{m-1}_{v=0} \biggl(\sum_{i\in Z_n}
\bigl[(t^{-i^2-3v+2v(i-v)})/(v)_q!\bigr]\Lba_{i-v}\biggr)a^{2v}e^vf^v\\
&= \sum^{m-1}_{v=0} \biggl(\sum_{i\in
Z_n}\bigl[(t^{-(i^2-2iv+v^2)-3v-v^2})/(v)_q!\bigr]\Lba_{i-v}\biggr)a^{2v}e^vf^v\
&=\sum^{m-1}_{v=0}\biggl(\sum_{i\in Z_n}
\bigl[(t^{-(i-v)^2-3v-v^2})/(v)_q!\bigr]\Lba_{i-v}\biggr)a^{2v}e^vf^v\\
&= \sum^{m-1}_{v=0} \biggl(\sum_{i\in Z_n}
\bigl[(t^{-3v-v^2})/(v)_q!\bigr]t^{-i^2}\Lba_i\biggr)a^{2v}e^vf^v.\endalign$$
Thus
$u=c\biggl(\sum\limits^{m-1}_{v=0}\bigl[(t^{-3v-v^2})/(v)_q!\bigr]
a^{2v}e^vf^v\biggr)$.\hfill$Q.E.D.$
\vskip2em
\sub{The Special Case $n=8$}
\mbk
Let $n=8$. Then $m=2$, $q=\sqrt{-1}$ and the algebraic relations for
$U_q\bigl(sl(2)\bigr)'$ are
$$\align
t^8 &=1, q=t^2\\
ae &=qea\\
af &= q^{-1}fa\\
a^8 &=1\\
e^2 &=0=f^2\\
[f,e] &= a^2-a^{-2}.\endalign$$
Note that by the previous calculation,
$$u=c(1+t^{-4}a^2ef)=c(1-a^2ef)$$
with $c$ given as in Lemma 3.

Recall that $\lba=\ol{a^{2(m-1)}e^{m-1} f^{m-1}}$ is a right integral
for $U_q(sl_2)'$. Thus, when $n=8$, the right integral is
$\lba=\ol{a^2ef}$.

\sub{Lemma 4} Let $X=-a^2ef$. Then $u=c(1+X)$ and
$X^2=(a^4-1)X=-2\biggl(\sum\limits_{i \te{ odd}} \Lba_i\biggr)X$.

\sub{Proof} $\qquad\qquad efef =
e[f,e]f+e^2f^2=e[f,e]f=e(a^2-a^{-2})f$\hfill\mbk\noi
$\ph{Proof \qquad\qquad efefef }
=(q^{-2}a^2-q^2a^{-2})ef=(a^{-2}-a^2)ef$.
\bbk\noi
Therefore $X^2=a^4(ef)^2=a^2(a^{-2}-a^2)a^2ef=(1-a^4)a^2ef=(a^4-1)X$.

Since $a^4=\sum\limits_{i\in Z_8} t^{-4i}\Lba_i=\sum\limits_{i\in Z_8}
(-1)^i\Lba_i$, we have
$$a^4-1=-2(\Lba_i+\Lba_3+\Lba_5+\Lba_7).$$
Thus
$$X^2=-2\biggl(\sum_{i \te{ odd}} \Lba_i\biggr)X.$$\hfill$Q.E.D.$
\vskip2em
The special grouplike in this case is $G=a^2$. Thus the ribbon element is
$v=G^{-1}u=a^2u$. Thus $v=a^2c(1+X)$.

To evaluate $\lba(v^k)$, let $H=(a)$ be the cyclic group generated by
$a$.  \hfill \break \noi Note that $v^k=c_0+c_1X$, where $c_i\in k[H]$.

\sub{Lemma 5} Writing $c_1=\sum\limits_{i\in Z_8} \alp_i\Lba_i$, with
$\alp_i\in k$, then $\lba(v^k)=(-1/8)\sum\limits_{i\in Z_8} \alp_i$.

\sub{Proof}
$$\align
c_1X &=\biggl(\sum_i\alp_i\Lba_i\biggr)
(-a^2ef)=-\biggl(\sum_i\alp_i\Lba_ia^2ef\biggr)=-\biggl(\sum_i\alp_i
\Lba_it^{-2i}\biggr) ef\\
&(a^2=\sum_{i\in Z_8} t^{-2i}\Lba_i \ \te{ by Lemma 2.}).\endalign$$
Thus $c_iX=- \sum\limits_{i,j\in Z_8}
\alp_i(t^{ij}/8)a^jt^{-2i}ef$.
$$\lba(v^k)=\ol{a^2ef}[c_1X]=- \sum_{i\in Z_8}
\alp_i(t^{2i}/8)t^{-2i}=-(1/8)\sum_{i\in Z_8} \alp_i.$$\hfill$Q.E.D.$
\vskip2em
\sub{Lemma 6} Let $n=8$ and let $\lba$ be the right integral and $v$ be
the ribbon element for $U_q\bigl(sl(2)\bigr)'$ as described above. Then
$\lba(v^k)=-k/2$.

\sub{Proof} $v^k=a^{2k}c^k(1+X)^k=\biggl(\sum\limits_{i}
t^{-2ik-i^2k}\Lba_i\biggr)(1+X)^k$, by Lemma 2. \hfill \break \noi Thus
$v^k=t^k\biggl(\sum\limits_it^{-k(i+1)^2}\Lba_i\biggr)(1+X)^k$.

Now assume that $k\ge 2$. Then $$(1+X)^k=1+kX+\sum^k_{j=2}{k\choose j}
X^j.$$
Since $X^2=-2(\Lba_1+\Lba_3+\Lba_5+\Lba_7)X$ by Lemma 4, it follows that
$$X^j=(-2)^{j-1}(\Lba_i+\Lba_3+\Lba_5+\Lba_7)X$$
for $j\ge 2$.

Thus
$$\align
(1+X)^k &=1+kX+\biggl(\sum^k_{j=2}{k\choose j}
(-2)^{j-1}(\Lba_1+\Lba_3+\Lba_5+\Lba_7)\biggr)X\\
&= 1+kX-(1/2)(\Lba_1+\Lba_3+\Lba_5+\Lba_7)\biggl(\sum^k_{j=2} {k\choose
j} (-2)^j\biggr)X\\
&=1+kX-(1/2)(\Lba_1+\Lba_3+\Lba_5+\Lba_7)\biggl(\sum^k_{j=0}{k\choose j}
(-2)^j-1+2k\biggr)X.\endalign$$
So
$$(1+X)^k=1+kX-(1/2)(\Lba_1+\Lba_3+\Lba_5+\Lba_7)\biggl(\sum^k_{j=0}{k\choose
j} (-2)^j-1+2k\biggr)X$$
for $k\ge 1$.

Since $v^k=t^k\biggl(\sum\limits_i t^{-k(i+1)^2}\Lba_i\biggr)(1+X)^k$,
the coefficient $c_1$ of $X$ in the expression \hfill
\break \noi$v^k=c_0+c_1X$ is:
$$\biggl(\bigl(t^k\bigl[\big(\sum_ikt^{-k(i+1)^2}\Lba_i\bigr)-(1/2)\bigl(\sum_{i
odd}} \bigl((-1)^k-1+2k\bigr)
t^{-k(i+1)^2}\Lba_i\bigr)\bigr]\bigr)\biggr).$$
Thus
$$\align
\lba(v^k) &= (-1/8)t^k\biggl(\sum_i
kt^{-k(i+1)^2}-(1/2)\biggl(\sum_{i\te {odd}}
\bigl((-1)^k-1+2k\bigr)t^{-k(i+1)^2}\biggr)\biggr)\\
&= (-1/8) t^k\biggl(\sum_{i \te{ even}} kt^{-k(i+1)^2}
-(1/2)\bigl((-1)^k-1\bigr)\biggl(\sum_{i\te{ odd}}
t^{-k(i+1)^2}\biggr)\biggr)\\
&= (-1/8)t^k\biggl(\sum_{i\te{ odd}}
kt^{-ki^2}-(1/2)\bigl((-1)^k-1\bigr)\biggl(\sum_{i\te{ even}}
t^{-ki^2}\biggr)\biggr).\endalign$$
Now $$\align
\sum\limits_{i\te{ odd}} kt^{-ki^2} &=
k(t^{-k}+t^{-9k}+t^{-25k}+t^{-49k})= k(4t^{-k}),\\
\sum_{i \te{ even}} t^{-ki^2} &= 1+t^{-4k}+
t^{-16k}+t^{-36k}=1+(-1)^k+1+(-1)^k\\
&=2\big(1+(-1)^k\bigr).\endalign$$

Hence
$$\align
\lba(v^k) &=(-1/8)t^k\bigl(4kt^{-k}-(1/2)\bigl(
(-1)^k-1\bigr)(2)\bigl(1+(-1)^k\bigr)\bigr)\\
&=(-1/8)\bigl(4k-\bigl((-1)^{2k}-1\bigr)\bigr)=-4k/8.\endalign$$
Thus $\lba(v^k)=-k/2.$\hfill$Q.E.D.$
\vskip2em
\sub{Corollary} The value of the 3-manifold invariant
$INV\bigl(L(k,1)\bigr)$ for $n=8$ is given by formula
$INV\bigl(L(k,1)\bigr)=\sqrt{-1} k$ for $k\ne 0$.

\sub{Proof} The surgery datum for $L(k,1)$ is an unknotted loop with $k$
curls.  Hence the unnormalized invariant is given by the formula
$TR(v^kG^{-1})=\lba(Gv^kG^{-1})=\lba(v^kG^{-1}G)=\lba(v^k)=-k/2$. The
normalized invariant is given by the formula
$$INV\bigl(L(k,1)\bigr)=\bigl[\lba(v)\lba(v^{-1})\bigr]^{-c(K)/2}\bigl[
\lba(v)/\lba(v^{-1})\bigr]^{-\sig(K)/2}TR(K).$$
Here $c(K)=1$ and $\sig(K)=1$ if $k>0$, $\sig(K)=-1$ if $k<0$ since the
link has one component, and the linking matrix is $(k)$.  We know that
$\lba(v)=-1/2$ and $\lba(v^{-1})=1/2$. Therefore
$$\align
INV\bigl(L(k,1)\bigr) &=
\bigl[(1/2)(-1/2)\bigr]^{-1/2}\bigl[(1/2)/(-1/2)\bigr]^{\pm 1}(-k/2)\\
&= (-2^2)^{1/2}(-1)(-k/2)=(-1)^{1/2}k.\endalign$$
This completes the proof.\hfill$Q.E.D.$
\vskip2em
\sub{Remark} This finishes our verification that the invariant $INV$ is
definitely different from the $WRT$ invariant in the case $n=8$, where
$WRT$ is trivial.  During the preparation of this paper, it has come to
our attention that similar results have been independently obtained by
Tomotada Ohtsuki [14].  He finds that invariants defined for
$U_q(sl_2)$' in a manner equivalent to ours necessarily vanish for
3-manifolds that are not rational homology spheres, and he performs
calculations similar to ours for Lens spaces.  It remains to be seen
whether these invariants see information beyond the homology for
rational homology spheres.  We are in the process of applying our
techniques to other finite-dimensional Hopf algebras (See [17]).  These
results will be discussed in a subsequent paper.
\vskip2em\noi
{\bf VI. The Quantum Double of a Finite Group}
\mbk
We now consider the 3-manifold invariant that is associated with the
quantum double of a finite group $G$.  This Hopf algebra will be denoted
by the notation $D(G)$.

\sub{Comment}  Let $k$ be a field.  The Hopf algebra described below is
the Drinfeld double of the Hopf algebra $F(G)$ where $F(G)$ denotes the
algebra of $k$-valued functions on $G$.

As a set, $D(G) = F(G) \otimes k[G]$ is the group ring of $G$ over $k$.
$D(G)$ has a basis, as a vector space over $k$, consisting in the
elements $(g,h) \epsilon G \times G$ where $(g,h)$ denotes
\hfill\break \noi $\delta_g
\otimes h \epsilon F(G) \otimes k[G]$.  Here $\delta_g$ is the
characteristic function $\delta_g:G \rightarrow k$ defined so that
\hfill\break \noi $\delta_g(g) = 1$ and $\delta_g(x) = 0$ if $x \ne g$.

We shall sometimes write $\delta_{g,x} = \delta_g(x)$.  The letter $e$
will denote the identity element in $G$.  The symbol 1 will denote the
identity element in $F(G)$.  Note that $$1 = \sum_{g \epsilon
G}\delta_ g.$$

Multiplication in $D(G)$ is given by the formula $$(g,h)(x,y) =
\delta_{h^{-1}gh, x}  (g,hy).$$

The coproduct in $D(G)$ is given by the next formula
$$\Delta(g,h) = \sum_{xy= g} (x,h) \otimes (y,h).$$

The sum is over all $x$ and $y$ in $G$ such that $xy = g$.

The counit is given by $E(g,h) = \delta_{g,e^.}$

The antipode is given by $s(g,h) = (h^{-1}g^{-1}h, h^{-1})$.

The braiding (Yang-Baxter) element in $A \otimes A$ has the formula
$$\rho = \sum_{g \epsilon G}(g,e) \otimes (1,g).$$

Observe that the Drinfeld element $u$ with $s^2(x) = u^{-1}xu$ for all
$x$ $\epsilon$ $D(G)$ has the formula $$u = \sum_{g\epsilon G} s
(1,g)(g,e) = \sum_{g \epsilon G} (g,g^{-1}).$$

It follows directly from the definition of $D(G)$ that $s^2(x) = x$ for
all $x$ in $D(G)$.  Thus $u$ is a central element in $D(G)$, $s(u) = u$
and $E(u) = 1$.  The special grouplike element in this Hopf algebra is
the identity element (1,e) and therefore $v = u$ is a ribbon element.

A right integral is given by the formula $\lambda (g,h) =
\delta_{e,h^.}$  In this case, the 3-manifold invariant that results
from the right integral is identical to the invariant obtained via a
matrix trace associated with the left regular representation of the
algebra on itself.  The proof of this statement will be given below, but
first it is worthwhile detailing the left regular representation.

Note that in $A$, we have the following products
$$\align
(1,h)(x,y) &= (hxh^{-1}, hy) \\
(g,h)(1,y) &= (g, hy).\endalign
$$
Consider the action of $R$ on $(D(G)$ that is obtained by left
multiplication:
$$\align
\rho\bigl[(a,b)\otimes (c,d)\bigr] &= \rho(a,b)\otimes (c,d)\\
&=\biggl(\sum_{g\in G} (g,e)\otimes (l,g)\biggr) (a,b) \otimes (c,d) =
\sum_{g\in G} (g,e)(a,b)\otimes (l,g)(c,d)\\
&=\sum_{g\in G} \dta_{g,a} (g,b)\otimes (gcg^{-1}, gd)=(a,b)\otimes
(aca^{-1}, ad).\endalign$$
Thus
$$\rho(a,b)\otimes (c,d) = (a,b)\otimes (aca^{-1}, ad).\tag{$*$}$$
Similarly,
$$\rho^{-1}(a,b)\otimes (c,d)=(a,b)\otimes (a^{-1}ca,
a^{-1}d).\tag{$**$}$$
The representation of $\rho$ maps $V \otimes V \rightarrow V \otimes V$;
$V$ denotes $D(G)$ as a vector space over the field $k$.  Since the
elements $(g,h) \epsilon G \times G$ index a basis for $V$, the
equations
$(*)$ and $(**)$ above define the matrices of $\rho$ and $\rho^{-1}$ in
this basis.

In order to define a link invariant via linear algebra, we need to first
define $R = \rho P$ and $\bar R = P \rho^{-1}$, where $P:V \otimes V
\rightarrow V \otimes V$ is the transposition, $P(x \otimes y) = y
\otimes x$.  Since $\rho$ satisfies the algebraic Yang-Baxter equation,
$R$ satisfies the knot theoretic braiding equation.

Along with $R$, we need the cup and cap mappings $\cup:k \rightarrow V
\otimes V$ and \hfill \break \noi $\cap: V \otimes V \rightarrow k$
satisfying the identities
demanded by Figure 1 (See the discussion in section 3).  In fact, if the
cup and cap maps are chosen properly, they can satisfy the properties
that correspond to the enlargement category of the Hopf algebra with
respect to the left regular representation.  That is, we would like to
choose cup and cap so that, with respect to this representation, the
rule for sliding an element around a maximum or minimum involving the
antipode is a fact of composition of mappings.  More precisely, let
$r:D(G) \rightarrow$ End(V) denote the left regular representation:
$r(x)y = xy$.  Then the antipode rule that is desired becomes
$(r(x) \otimes 1) \cup = (1 \otimes r(s(x))) \cup$ and $\cap (1 \otimes
r(x)) = \cap (1 \otimes r(s(x)) \otimes 1)$ for all $x$ in $D(G)$.
\mbk
{\bf Lemma:}  With respect to the basis $\{(g,h)\}$ for $D(G)$ the
equation
$\cap (1 \otimes r(x)) = \cap (r(s(x)) \otimes 1)$ for all $x$ in $D(G)$
becomes $\cap [(a,b) \otimes \delta_{h^{-1}gh, c}(g,hd)] \hfill \break
\noi= \cap [\delta_{g,a}(h^{-1}gh, h^{-1}b) \otimes (c,d)]$.
A similar result holds for the corresponding equation for the cup.

{\bf Proof.}
$$\align
\cap (1 \otimes r(g,h))[(a,b) \otimes (c,d)] &= \cap[(a,b)
\otimes (g,h)(c,d)]\\ &= \cap [(a,b) \otimes \delta_{h^{-1}gh,
c}(g,hd)].\endalign$$
$$\align
\cap (r(s(g,h)) \otimes 1)[(a,b) \otimes (c,d)]
&= \cap [s(g,h)(a,b) \otimes (c,d)]\\
&= \cap [(h^{-1}gh,h^{-1})(a,b) \otimes (c,d)]\\
&= \cap [\delta_{g,a}(h^{-1}gh,h^{-1}b) \otimes (c,d)].\endalign$$
This completes the proof.\hfill$Q.E.D.$
\vskip2em
We see at once from this Lemma that the following definitions of cup and
cap satisfy the compatibility with the antipode:
$$\cap(a,b) \otimes (c,d) = \delta_{a,c^{-1}} \delta_{b,d}$$
$$\cup (1) = \sum_{a,b \epsilon G} (a,b) \otimes (a^{-1},b)$$.
\vfill\eject
This definition also satisfies the required compatibility with $R$ and
$\bar R$.  For example, view the diagrams below.  Here we indicate only
the possible non-zero terms for $R$, $\bar R$ and their compatibility
with the switching of an arc across a maximum.  We omit the verification
of the remaining identities.

\vskip25em
We are now in a position to examine the linear algebraic formulation of
this invariant in two ways.  We can apply the linear algebra directly,
or we can slide the elements of the Hopf algebra on the diagram,
concentrating them in one vertical arc and apply the trace associated
with the cup and the cap.  Given a knot or link $K$, let $\lgl K\rgl$
denote
the evaluation produced by the linear algebraic model.  Suppose that an
element $w$ of $D(G)$ is concentrated in the left vertical portion of an
unknotted circle as shown below.

\vskip5em \noi
Call this curve with algebra attached $O(w)$.  Then we see that $<O(w)>
= \cap (w \otimes 1) \cup (1)$ where 1 is the identity element in $k$.
Now the 1 in $w \otimes 1$ is the identity element in $D(G)$.  Hence
this $ 1 = \sum_{g \in G} (g,e)$.  Thus
$$\align
\cap\bigl(a,b)\otimes 1\bigr)\cup(1) &=\cap\bigl((a,b)\otimes 1\bigr)
\sum_{g,h\in G} (g,h)\otimes (g^{-1},h)\\
&=\cap\biggl\{\sum_{g,h\in G} (a,b)(g,h)\otimes (g^{-1},h)\biggr\}\\
&=\cap\biggl\{\sum_{g,h\in G}\dta_{b^{-1}ab,g}(a,bh)\otimes
(g^{-1},h)\biggr\}\\
&=\sum_{g,h\in G} \dta_{b^{-1}ab,g}\cap\bigl\{(a,bh)\otimes
(g^{-1},h)\bigr\}\\
&=\sum_{g,h\in G}\dta_{b^{-1}ab,g}\dta_{a,g}\dta_{bh,h}\\
&=|G|\dta_{b,e}=|G|\lba(a,b)\endalign$$
where $\lambda$ denotes the right integral on $D(G)$, as we described it
at the beginning of the section.  It follows that {\it the invariant}
$|G|^{-c(K)}\lgl K\rgl$ {\it defined by linear algebra is identical to
the invariant defined on the Hopf algebra by the right integral}.
\bbk\noi
{\bf Theorem}.  Let $M^3(K)$ denote the 3-manifold obtained by doing
surgery on the link $K$ given in the blackboard framing.  Let
$\pi_1(M^3(K))$ denote the fundamental group of the manifold $M^3(K).$
Then the invariant $\lgl K\rgl$ is equal to the number of homomorphisms of
$\pi_1(M^3(K))$ to the finite group $G$. Hence the invariant defined
via the right integral on the quantum double of the finite group also
counts these homomorphisms.
\bbk\noi
{\bf Proof}.  It follows from our description of the linear algebra that
the value of $\lgl K\rgl$ is equal to the number of distinct labellings
of
the diagram with basis elements $(g,h) \epsilon G \times G$ according to
the
following rules:  The diagram is regarded as a composition of fragments
corresponding to $R$ $\bar R$, $\cup$, $\cap$.  The extreme of these
fragments are nodes labelled with elements of $G\times G$, and they
satisfy the diagrammatic rules shown below:
\vskip12em
These rules mean that each such labelling of the diagram gives a
homomorphism of the group that is defined via generators and relations
corresponding to exactly these rules in the context of the given
diagram.  This abstract group is precisely the fundamental group of the
3-manifold obtained by surgery on the link.  The relations derived from
the identities in the first coordinates define the fundamental group of
the complement of the knot.  The identity in the second coordinates
insures that the framing longitude is set equal to the identity.  This
describes the fundamental group of the 3-manifold.\hfill$Q.E.D.$
\vskip2em
\sub{Remark}  Since the presentation of the abstract group associated
with the diagram as described in the above proof may look unfamiliar to
the reader, we note that the first coordinate relations do indeed
present the fundamental group of the link complement.  This presentation
is a presentation similar to the classical Wirtinger presentation, but
formulated with respect to an unoriented diagram arranged with respect
to a vertical direction.  In this formulation, the correspondence
between labelling conventions and the relations in the fundamental group
are shown in Figure 2.

In this presentation of the fundamental group of the link complement
each generator of the fundamental group is given an orientation so that
its linking number with a vertical segment passing through it is equal
to $-1$.  All the relations are expressed in the language of this system
of generators.  Since the linking number of a generator with a vertical
segment changes when the generator is homotoped around a maximum or a
minimum in the link diagram, we need to also include the relations at
the maxima and minima as shown below.
\vskip8em
In Figure 2, we have illustrated this relation by showing the loops $a$
and $a^{-1}$ on either side of a maximum.
\vfill
\centerline {\bf Figure 2}
\eject
For example, here is the diagram and reduction of the relations for the
fundamental
group of the complement of the trefoil knot in this unoriented
formalism.
\vskip20em
Note that this system of relations is exactly what is demanded by the
first coordinates of the labelling for states of the invariant $\lgl K\rgl$.
The second coordinate relation is equivalent to saying that $\mu = e$
where $\mu$ is the product of the generators of the fundamental group
that label overcrossing lines encountered in one trip around the diagram
(for each component).  This element $\mu$ is the image in the
fundamental group of the framing longitude for the surgery.  It
therefore follows that the full set of relations (fundamental group plus
$\mu = e$ for each component) describes the fundamental group of the
3-manifold obtained by surgery on the link.  These remarks complete the
proof of the Theorem.

\sub {\bf Remark}  The main theorem in this section has been conjectured
(and presumably proved) for a representation theoretic version of the
invariant associated with the quantum double of a finite group by D.
Altschuler and A. Coste [1 and 1.1].  Our argument verified directly
that this result hold for the corresponding invariant using the right
integral or, equivalently, the left regular representation of the
quantum double.  Other formulations of this invariant in the context of
topological quantum field theory may be found in [2], [4].

\sub {\bf Remark}  It is interesting to view examples of this invariant,
calculated in the Hopf algebra context.  Sliding the algebra around the
diagram yields a sum of products in $D(G)$ that resolves into
restrictions that correspond to the desired group representation, but
without the guidance of the linear algebraic model, the relation with
the fundamental group of the 3-manifold is not obvious even though it
can be easily verified in small cases.
\vfill\eject
\heading REFERENCES\endheading
\baselineskip=12pt
\frenchspacing
\bbk
\item{( 1.)} {\smc Altschuler, D. and Coste, A.\/} Invariants of
3-manifolds from finite groups.  In {\it Proceedings of the XXth
International Conference on Differential Methods in Theoretical Physics}
- June 1991 - N.Y., N.Y., World Sci. Pub.
\mbk
\item{(1.1)} {\smc Altschuler, D. and Coste, A.\/} Quasi-quantum
groups,
knots, three-manifolds, and topological field theory.  Commun. Math.
Phys. {\bf 150} (1992), 83-107.
\mbk
\item{( 2.)} {\smc Dijkgraaf, R. and Witten, E.\/} Topological gauge
theories and group cohomology. Commun. Math. Phys. {\bf 129} (1990),
393-429.
\mbk
\item{( 3.)} {\smc Drinfeld, V. G.\/} Quantum groups. {\it Proceedings
of
the International Congress of Mathematicians}, Berkeley, California, USA
(1987), 798-820.
\mbk
\item{( 4.)} {\smc Freed, D. S. and Quinn, F.\/} Chern-Simons theory
with finite gauge group.  Commun. Math. Phys. (to appear).
\mbk
\item{( 5.)} {\smc Hennings, M. A.\/} Invariants of links and
3-manifolds obtained from Hopf algebras. (preprint 1989).
\mbk
\item{( 6.)} {\smc Kauffman, L. H.\/} From knots to quantum groups and
back.  {\bf In Quantum Groups} (edited by Fairlie and Zachos, World Sci.
1991.), 1-33.
\mbk
\item{( 7.)} {\smc Kauffman, L. H.\/} Gauss Codes, Quantum Groups and
Ribbon Hopf Algebras.  Reviews in Mathematical Physics, Vol.5, No. 4
(1993), 735-773.
\mbk
\item{( 8.)} {\smc Kauffman, L. H.\/} {\it Knots and Physics}. World
Sci. Pub. (1991), (2nd edition 1993).
\mbk
\item{( 9.)} {\smc Kauffman, L. H. and Radford, D. E.\/} A necessary and
sufficient condition for a finite-dimensional Drinfel'd double to be a
ribbon Hopf algebra.  Journal of Algebra.  Vol. 159, No. 1, August 1,
(1993), pp. 98-114.
\mbk
\item{(10.)} {\smc Kauffman, L. H. and Radford, D. E.\/} Hopf Algebras
and Invariants of 3-Manifolds.  (Announcements - July 1992, July 1993).
\mbk
\item{(11.)} {\smc Kirby, R. and Melvin, P.\/} The 3-manifold invariants
of Witten and Reshetikhin-Turaev for sl(2,C).  Invent. Math. 105,
473-545 (1991).
\mbk
\item{(12.)} {\smc Larson, R. G. and Sweedler, M. E.\/} An associative
orthogonal bilinear form for Hopf algebras.  Amer. J. Math {\bf 91}
(1969), 75-94.
\mbk
\item{(13.)} {\smc Lawrence, R. J.\/} A universal link invariant using
quantum groups.  {\it Differential Geometric Methods in Theoretical
Physics}, Chester (1988), {\bf 55-63}.
\mbk
\item{(14.)} {\smc Ohtsuki, T.\/} Invariants of 3-manifolds derived from
universal invariants of framed links.  (preprint 1993).
\mbk
\item{(15.)} {\smc Radford, D. E.\/} The trace function and Hopf
algebras.  (to appear in Journal of Algebra).
\mbk
\item{(16.)} {\smc Radford, D. E.\/} Generalized double crossproducts
associated with quantized enveloping algebras.  (preprint).
\mbk
\item{(17.)} {\smc Radford, D. E.\/} On Kauffman's knot invariants
arising from finite dimensional Hopf algebras, {\it Advances in Hopf
algebras,\/
applied mathematics, Marcel Dekker, N.Y., 1994.
\mbk
\item{(18.)} {\smc Reshetikhin, N. Yu. and Turaev, V. G.\/} Ribbon
graphs
and their invariants derived from quantum groups.  Commun. Math Phys.
127, 1-26 (1990).
\mbk
\item{(19.)}  {\smc Reshetikhin, N. Yu. and Turaev, V. G.\/} Invariants
of 3-manifolds via link polynomials and quantum groups.  Invent. Math.
{\bf 103}, 547-597  (1991).
\mbk
\item{(20.)} {\smc Sweedler, M. E.\/} {\bf Hopf Algebras}.  Mathematics
Lecture Notes Series, Benjamin, New York, 1969.
\mbk
\item{(21.)} {\smc Witten, E.\/} Quantum field theory and the Jones
polynomial.  Commun. Math. Phys. 121 (1989), 351-399.
\vfill\eject\bye